          [2001/04/25 1.1 (PWD)]
\documentclass[a4paper,twocolumn]{esapub} 
\usepackage{natbib}
\usepackage{psfig}
\usepackage{graphicx}

\title{Scientific requirements of ALMA, and its capabilities for key-projects: Extragalactic}
\author{Carlos De Breuck}
\affil{European Southern Observatory, Karl Schwarzschild Stra\ss e 2, D-85748 Garching, Germany}

\newcommand{\farcs}{\ifmmode \rlap.{^{\prime\prime}}\else
    $\rlap.{^{\prime\prime}}$\fi}
\newcommand{\degr}{\ifmmode {^{\circ}}\else {$^\circ$}\fi}
\newcommand{\eg}{{\it e.g.} }
\newcommand{\etal}{{\it et al.~}}

\begin{document}


\maketitle

\begin{abstract}
The Atacama Large Millimeter Array (ALMA) consists of 64 antennas of 12\,m diameter that will initially observe in 4 frequency bands between 84 and 720 GHz with spatial resolutions down to 0\farcs01 and velocity resolutions as fine as 0.05 km/s. These technical requirements are based on three primary science goals. We illustrate two of these requirements: (i) the ability to detect spectral line emission from a Milky-Way type galaxy at $z$=3, and (ii) the ability to provide precise images at an angular resolution of 0\farcs1. 
Finally, we present a possible large extragalactic project with ALMA: molecular line studies of submm galaxies.
\end{abstract}

\section{Primary Scientific Requirements}

ALMA will be a flexible observatory supporting a breath of research in the fields of planetary, galactic and extra-galactic astronomy. The three primary science requirements have been defined as:
\begin{enumerate}
\item The ability to detect spectral line emission from CO or C{\small I} in a normal galaxy like the Milky Way at a redshift of $z=3$, in less than 24 hours of observation.
\item The ability to image the gas kinematics in protostars and protoplanetary disks around young Sun-like stars at a distance of 150~pc, enabling one to study their physical, chemical and magnetic field structures and to detect the gaps created by planets undergoing formation in the disks.
\item The ability to provide precise images at an angular resolution of 0\farcs1. Here, the term precise images means representing within the noise level the sky brightness at all points where the brightness is greater than 0.1\% of the peak image brightness. This requirement applies to all sources visible to ALMA that transit at an elevation greater than 20\degr.
\end{enumerate}

We now concentrate on the two requirements important for extragalactic astronomy (see J. Richer's contribution for the galactic requirements).

\subsection{Spectral line imaging of normal galaxies at $z$=3}
An estimate of the technical requirements to achieve this science goal can be made by using experience learned from existing millimeter arrays, which have collecting areas between 500 and 1000~m$^2$. These arrays now routinely detect CO emission from high redshift galaxies and quasars \citep[see ][for a review]{car04a}. Fainter molecular lines such as HCN \citep[\eg][]{car04b} and atomic Carbon \citep{wei04a} are now also within the reach of existing telescopes. These observations take one to two days of total observing time, and are only possible for the most luminous sources and/or with the aid of gravitational lensing. In normal, unlensed galaxies, these lines would be a factor of 20 to 30 fainter.

The sensitivity of ALMA for a given integration time is essentially controlled by three major terms: (1) the atmospheric transparency, (2) the noise performance of the detectors, and (3) the total collecting area. 

The location of ALMA on the Chajnantor plateau at an altitude of 5000m will minimize the contribution from the atmosphere compared to existing millimeter observatories. The noise level of the detectors can be reduced be a factor of two, and will then approach the fundamental quantum limit. An important factor of $\sqrt 2$ will be gained by the requirement that ALMA support front end instrumentation capable of measuring both states of polarization. The remaining factor of 7 to 10 can only be gained by increasing the collecting area. Hence, an ALMA requirement is a collecting area $>$7000~m$^2$.
\vspace{1.5cm}

A more specific calculation of the requirement to have the ability to detect CO emission from a Milky Way type galaxy at $z$=3 is as follows. At cosmological distances, the 10\,kpc disk of the Milky Way is much smaller than the primary beam of existing millimeter antennas, so a single observation would be sufficient. The flux density sensitivity in an image from an interferometric array can be written as $$\Delta S = \frac{4 \sqrt 2 k T_{\rm sys}}{\gamma \epsilon_q \epsilon_a \pi D^2 \sqrt{n_p \frac{N(N-1)}{2}\Delta\nu \Delta t}}~{\rm W m^{-2} Hz^{-1},}$$ where $T_{\rm sys}$ is the system temperature, $\epsilon_a$ is the aperture efficiency, $\epsilon_q$ is the correlator quantization efficiency, $D$ is the antenna diameter, $n_{\rm p}$ is the number of simultaneously sampled polarizations, $N$ is the number of antennas, $\Delta \nu$ is the bandwidth, $\Delta t$ is the integration time, and $\gamma$ is a gridding parameter that we set to unity. For ALMA, we shall assume $\epsilon_q$=0.95, $n_{\rm p}$=2, and $\sqrt{N(N-1)}\simeq N$ so this equation then simplifies to $$\Delta S= \frac{2.6\times 10^6 T_{\rm sys}}{\epsilon_a N D^2\sqrt{\Delta\nu \Delta t}}~{\rm mJy}.$$

$T_{\rm sys}$ and $\epsilon_a$ vary between the different atmospheric bands of ALMA. A surface accuracy of 20\,$\mu$m should be achievable, providing aperture efficiencies between 0.75 and 0.45 (see Table~\ref{table:eff+Tsys}). $T_{\rm sys}$ depends on several atmospheric and instrumental parameters. Table~\ref{table:eff+Tsys} lists estimates of the achievable $T_{\rm sys}$ at an elevation of 50\degr\ using state-of-the-art receivers. With these assumptions, one can thus calculate the required collecting area parametrized by the product $ND^2$ to achieve a certain flux density level.

\begin{figure*}[ht]
\centering
\includegraphics[width=0.5\linewidth,angle=-90]{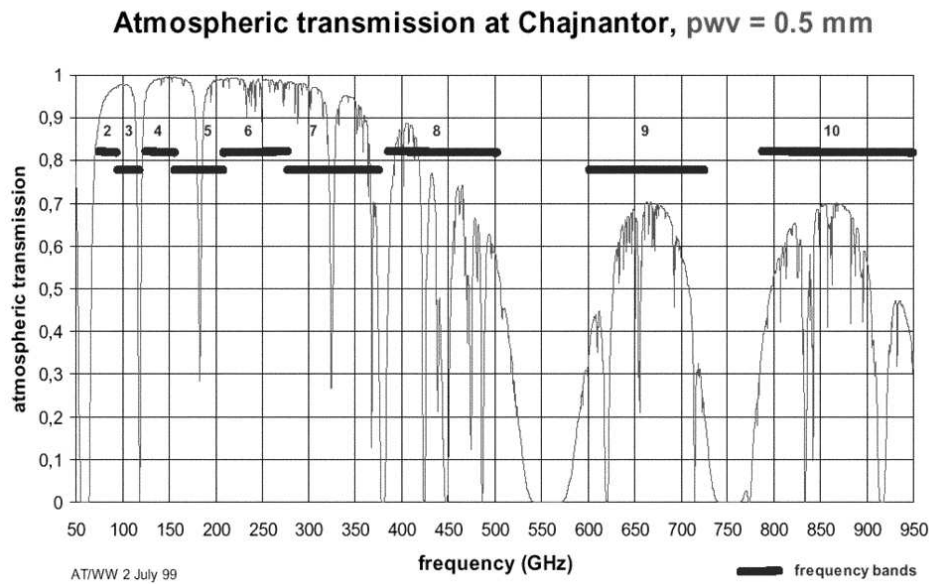}
\caption{Atmospheric transmission at Chajnantor with the ALMA frequency bands indicated. Initially, only bands 3, 6, 7 and 9 will be used .\label{fig:atmosphere}}
\end{figure*}

\begin{table}
  \begin{center}
    \caption{Aperture efficiencies $\epsilon_a$ and estimated system temperatures $T_{\rm sys}$ for ALMA at 50\degr elevation.}\vspace{1em}
    \renewcommand{\arraystretch}{1.2}
    \begin{tabular}[h]{ccc}
      \hline
      Frequency & $\epsilon_a$ & $T_{\rm sys}$ \\
      GHz & & K\\
      \hline
       35 & 0.75 &   35 \\
      110 & 0.74 &   67 \\
      230 & 0.72 &  107 \\
      345 & 0.69 &  251 \\
      409 & 0.67 &  425 \\
      675 & 0.54 & 1050 \\
      850 & 0.45 & 1150 \\
      \hline \\
      \end{tabular}
    \label{table:eff+Tsys}
  \end{center}
\end{table}

The total CO luminosity of the Milky Way in the CO(1--0) transition $L^{\prime}_{\rm CO(1-0)}$=3.7$\times$10$^8$\,K\,km\,s$^{-1}$pc$^2$ has been estimated by \citet{sol89}. The CO luminosities seen in higher CO transitions by COBE \citep{ben94,wri91} are slightly higher. In the following, we shall adopt $L^{\prime}_{\rm CO(1-0)}$=5$\times$10$^8$\,K\,km\,s$^{-1}$pc$^2$. From this, we can calculate the expected received flux density following \citet{sol92} as $$S_{\rm CO} = 3.08\times 10^{-8}\frac{L^{\prime}_{\rm CO}\nu_{\rm rest}^2(1+z)}{\Delta v_{\rm rest}D_L^2},$$ where $S_{\rm CO}$ is the flux density in Jy, $\nu_{\rm rest}$ is the rest frequency of the transition in GHz, $D_L$ is the luminosity distance in Mpc, and $\Delta v_{\rm rest}$ is the rest line width in km\,s$^{-1}$.

In the Milky Way, most of the CO emission arises in clouds of several tens of Kelvin kinetic temperature. In Milky Way-like galaxies at $z$=3, the gas will be somewhat warmer due to the higher background temperature from the cosmic microwave background radiation. At $z$=3, we can observe the J=3--2 or J=4--3 transitions, which fall inside the atmospheric transmission regions. We thus need to consider the expected luminosity of these higher order transitions, using a proper radiative transfer model that takes the higher background temperatures into account. Several such models have been proposed \citep{silk97,com99,pap00}. However, they reach significantly different conclusions.
There have also been a few observations of CO ladders in high redshift quasars \citep{car02,ber03} and Ultra Luminous InfraRed Galaxies \citep{wei04b}, all showing CO intensities up to the 6--5 or 7--6 transition. However, the CO emission from quasars and ULIRGs is likely to be dominated by the central component, while our requirement is to detect the outer regions in Milky Way type galaxies. In a sample of 28 nearby galaxies, \citet{mau99} finds $I_{\rm CO(3-2)}$/$I_{\rm CO(1-0)}$ ratios between 0.2 and 0.7, very different from the ones found in the high-$z$ quasars and ULIRGs. Given these uncertainties, we shall assume $L^{\prime}_{\rm CO(3-2)}$/$L^{\prime}_{\rm CO(1-0)}$=1.

For a standard $\Lambda$CDM cosmology with H$_0$=71\,km\,s$^{-1}$\,Mpc$^{-1}$,
$\Omega_{\rm M}$=0.27 and $\Omega_{\Lambda}$=0.73, the luminosity distance at $z$=3 is $D_L$=26\,Gpc. Assuming an intrinsic width $\Delta v$=300\,km$^{-1}$, the expected flux density of the CO(3--2) line is thus 36\,$\mu$Jy.
Requesting a 5$\sigma$ detection in a 75\,km\,s$^{-1}$ channel in 12 hours of on-source integration time (corresponding to 16 hours of total telescope time), we thus require an $ND^2$ of $\sim$7300\,m$^2$. This can be achieved with the ALMA array of 64 12\,m antennas. Of course, larger values of $N D^2$ are always desirable, as they would allow one to resolve the line flux density into more pixels (higher angular or spectral resolution) or image to higher S/N more quickly.

Next to CO, ALMA should also be able to observe other lines such as C{\small I}, N{\small II} and C{\small II} lines at cosmological redshifts. These lines will provide important probes of the IMF and the Lyman continuum luminosity from the most luminous stars in early galaxies. However, because the evolution of their luminosity as a function of redshift and Hubble type is less known than for CO, we did not use them to determine the total aperture requirement of the array.

\subsection{Precise high-resolution imaging}
The requirement for ALMA to obtain precise images at an angular resolution of 0\farcs1 follows from the need to complement contemporary facilities such as the James Webb Space Telescope, the extended VLA, and adoptive optics imaging on large ground based telescopes.

To obtain high fidelity images with an interferometer requires a sufficient number of baselines to adequately cover the $uv$ plane. To reach such excellent images limited by dynamic range requires that 50\% of the ($u,v$) cells be filled \citep{mor96}. This fraction of occupied cells (FOCC) is calculated out to the longest array baselines. The FOCC is a function of hour angle coverage. Obviously, one would like to observe sources within a limited hour angle range to avoid large system temperature variations that would corrupt the images. Especially in the submillimetre windows, such variation limit the hour angle coverage to approximately 2 hours. \citet{hol98} presents a detailed analysis of the variation of FOCC as a function of array configuration and hour angle coverage. To achieve an FOCC$>$0.5 in a configuration with a maximum baseline of 3000\,m in 2 hours of hour angle coverage requires a collecting length $ND>$560. This can be achieved with the ALMA array of 64 12\,m antennas.

As for any array, ALMA will also be prone to the short spacing problem. Because one cannot measure the smallest spatial frequencies, below approximately the antenna diameter, the interferometer will not be sensitive to sources more extended than $\sim$2/3 of the primary beam. A key requirement of ALMA is therefore the ability to observe in total power mode. To reach a similar S/N level in total power would require 4 antennas optimized for total power measurements (using a nutating secondary), each observing 4 times longer than the array (hence only 25\% of the projects will have total power information).

Even after the combination with the total power data, there will still be a gap in the $uv$ plane, located in a ring of approximately half the antenna diameter. This gap will be filled in by the Atacama Compact Array, a set of twelve 7\,m antennas.

\section{Detailed requirements of ALMA}
To achieve the above science requirements requires a reconfigurable array covering baselines from a few meters up to several kilometers, observing in all the millimeter and submillimeter windows (Fig.~\ref{fig:atmosphere}). The 12\,m diameter of the antennas is driven by the required pointing and surface accuracy. Additionally, the ALMA antennas will be equipped with water vapor radiometers to measure atmospheric pathlength variations. Together with a fast switching technique, this will minimize the image distortions caused by phase variations.

Given the diverse scientific community to be served by ALMA, the final major requirement of is that ALMA should be an easy by non-experts. Automated image processing will be developed and applied to most ALMA data, with expert help available for intricate experiments. Table~\ref{table:requirements} summarizes the requirements of ALMA.

\begin{table*}
  \begin{center}
    \caption{Requirements of ALMA}\vspace{1em}
    \renewcommand{\arraystretch}{1.2}
    \begin{tabular}[h]{|l|l|}
      \hline
      Requirement & Specification \\
      \hline
      Frequency & all atmospheric windows between 30 and 950\,GHz (Fig.~\ref{fig:atmosphere})\\
      Bands & 10 bands; initial priority to band 3=84--116\,GHz, band 6=211-275\,GHz, \\
      & band 7=275--373\,GHz and band 9=602--720\,GHz\\
      Tunability & possible completely across all observable windows\\
      Spectral resolution & sufficient (0.01 km/s) at 100\,GHz to resolve thermal line widths\\
      Intraband tuning & within 1.5\,s\\
      Interband tuning & within 1 minute; within 1.5\,s if standby\\
      Dynamic range & spectral: 10000:1; imaging: 50000:1\\
      \hline
      Flux sensitivity & sub-mJy point source at all frequencies within 10\,min\\
      & under median atmospheric conditions\\
      Site & Llano de Chajnantor at 5000\,m altitude\\
      Antennas & 64 antennas of 12\,m diameter\\
      Antenna surface & rms deviations of 25\,$\mu$m from ideal\\
      Receivers & close to quantum limited\\
      IF bandwidth & 8\,GHz per polarization in continuum mode\\
      Dynamic scheduling & optimization following scientific priority and required/current conditions\\
      \hline
      High fidelity & on spatial scales of degrees to 0\farcs01\\
      Total power & 4 antennas equipped with nutating subreflectors\\
      Configurations & continuous from within 150\,m to maximum baseline of 18.5\,km\\
      Pointing & accurate to 0\farcs6 using reference pointing\\
      Antenna locations & determined to 65\,$\mu$m\\
      Phase corrections & corrected phase visibility fluctuations not to exceed 1 radian at 950\,GHz\\
      Amplitude fluctuations & $<$3\% at 300\,GHz and $<$5\% at higher frequencies\\
      \hline
      Polarization & all polarization cross-products measured simultaneously\\
      Polarized flux error & $<$0.1\% of total intensity for $V$=0 source\\
      Polarization position angle & better than 6\degr\ for linear polarization\\
      \hline
      Calibration & accurate to 3\% below 300\,GHz and 5\% at higher frequencies.\\
      & Absolute calibration to 5\%.\\
      \hline
      Solar & it shall be possible to observe the Sun at all frequencies\\
      \hline
      Software & preparation, scheduling and reduction software provided by ALMA\\
      Data reduction & pipeline with minimal input from astronomer for most projects\\
      \hline
      \end{tabular}
    \label{table:requirements}
  \end{center}
\end{table*}

\section{Possible extragalactic programs}
In order provide a set of high-priority ALMA projects that could be carried out in $\sim$3--4 years of full ALMA operations, a Design Reference Science Plan (DRSP) has been set up. The DRSP contains 128 projects\footnote{The entire list of DRSP projects is available from http://www.eso.org/projects/alma/} written by 43 experts in four scientific categories (Galaxies and Cosmology,  Star and planet formation,  Stars and their evolution,  Solar system). These projects assume the full array of 64 antennas, which will be available by 2012. The scientific goals will of course evolve over time, so these projects may become slightly outdated by this time. The main goal of the DRSP is to serve as a quantitative reference for developing the science operations plan, for performing imaging simulations, and for software design. These provide a useful overview of ALMA's capabilities in these four domains. The DRSP does {\bf not} form the basis for any definition of ALMA early science observing programs, nor for any claims on key or other projects. Here we illustrate an example extragalactic project doing molecular line studies of submm galaxies.
\begin{figure}[ht]
\centering
\includegraphics[width=1.0\linewidth]{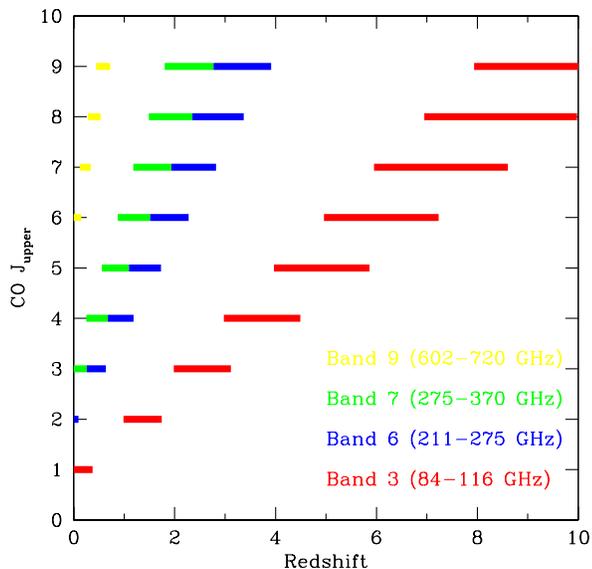}
\caption{The redshift coverage within the initial four ALMA frequency bands of the different rotational transitions of $^{12}$CO as a function of redshift. Note that at each redshift, there will be at least one CO line that can be detected with ALMA. At $z>3$, one can even detect two transitions within band~3.\label{fig:coladders}}
\end{figure}

The discovery of a significant population of dusty star forming galaxies at high redshift from deep (sub)mm surveys made with the SCUBA and MAMBO bolometer arrays has transformed our knowledge of galaxy formation \citep[\eg][]{sma97,gre04}. These sources make up at least half of the FIR/submm background \citep{hau98}.
A significant fraction (of order 50\%) of star formation in the cosmos occurs in these galaxies that are heavily obscured by dust. The optical/near-IR identification is extremely time-consuming, and often requires ultra-deep radio maps to narrow down the large positional uncertainty of the (sub)mm positions \citep{ivi02,dan04}. As a result, the redshift distribution is still not properly determined, although the median redshift is claimed to be close to $\langle z \rangle \sim 2.4$ \citep{cha03}.

ALMA will not only provide sub-arcsecond resolution images of these sources, solving the optical identification problem, but will also allow to bypass the optical spectroscopy altogether. The limited bandwidth of present-day mm interferometers means that accurate optical redshifts are needed before one can confirm the redshift by observing the CO lines \citep{neri03}. The 2$\times$4~GHz bandwidth of ALMA allows one to detect at least one CO transition in three frequency settings between 90 and 116\, GHz (Fig.~\ref{fig:coladders}). A second search will then be required to confirm the redshift. The two CO lines will also provide estimates of CO excitation conditions. For sources with 850\,$\mu$m continuum flux densities of 1\,mJy (which would be found with second generation bolometer arrays such as SCUBA-2/JCMT or LABOCA/APEX), ALMA would have a solid detection in $<$2\,hours integration time per source, so it is feasible to observe a representative sample of 50 sources in 100 hours. These sources can the be followed up with high resolution CO images to determine their composition (many sources are expected to be in mergers), and to derive dynamical mass estimates from their velocity profiles \citep[\eg][]{gen03}. Such studies require only 1 hour per source, compared to 24 hours or more for brighter sources known today.  Finally, the HCN line, which is typically 10$\times$ fainter than CO will be detectable in about 10 hours with ALMA. This will provide a much better tracer of the dense gas feeding the star-formation in these galaxies.

\section*{Acknowledgments}

Most of this paper is based on the draft ALMA Scientific Specification and Requirements in the version by Al Wooten and Tom Wilson, with important contributions from Bryan Butler.

\end{document}